\documentclass[sn-mathphys,Numbered]{sn-jnl}

\usepackage{graphicx}%
\usepackage{multirow}%
\usepackage{amsmath,amssymb,amsfonts}%
\usepackage{amsthm}%
\usepackage{mathrsfs}%
\usepackage[title]{appendix}%
\usepackage{xcolor}%
\usepackage{textcomp}%
\usepackage{manyfoot}%
\usepackage{booktabs}%
\usepackage{algorithm}%
\usepackage{algorithmicx}%
\usepackage{algpseudocode}%
\usepackage{listings}%
\usepackage{cleveref}

\begin{document}

\title[Dark Photons from displaced vertices]{Dark Photons from displaced vertices}


\author*[1]{\fnm{Triparno}
\sur{Bandyopadhyay}}\email{triparnb@srmist.edu.in}



\affil*[1]{\orgdiv{Department of Physics and Nanotechnology},
\orgname{College of Engineering and Technology, SRM Institute of Science
and Technology}, \orgaddress{\street{SRM Nagar}, \city{Kattankulathur},
\postcode{603203}, \state{Tamil Nadu}, \country{India}}}




\abstract{
    We investigate the unexplored regions of the dark photon parameter space to find
    a search strategy suitable to probe these. We show how displaced track searches
    at colliders with large 4pi trackers around the interaction point are excellent
    choices for exploring these uncharted regions. As an example, we study in detail
    the sensitivity of the Belle II trackers to dark photons which mediate interactions
    between the visible and dark sectors. We also show that the same strategy can
    be employed by other experiments to achieve the same goal.
}




\maketitle

\section{Introduction}\label{sec:sec1} With supersymmetry and exotica
searches at the Large Hadron Collider (LHC) experiments returning
nothing but exclusion plots for the last
decade~\cite{ParticleDataGroup:2022pth, Asai:2021ehn, Bose:2022obr,
Sekmen:2022vzu, Fox:2022tzz}, the interest of the particle physics
community has shifted substantially towards light, \(\ll 1\) GeV
particles. It is here that physics related to axion-like
particles~\cite{Georgi:1986df, Krauss:1987ud, Fukuda:2015ana,
Alves:2017avw, Bauer:2021wjo, Bandyopadhyay:2021wbb}, dark
photons (DP)~\cite{Okun:1982xi, Galison:1983pa, Holdom:1985ag},
$L_i-L_j$/ $B-L$ bosons~\cite{Fayet:1980rr, Foot:1990mn, He:1990pn,
He:1991qd}, the dark QCD spectrum~\cite{Cheng:2021kjg}, protophobic
bosons \cite{Feng:2016jff, Feng:2016ysn} etc have captured
the imagination of the community. Needless to say, this interest is not
spurious as these hypothetical particles are invoked as solutions to
many open questions in physics. These include, but are not limited to,
the mediation between the visible and the dark
sector~\cite{Boehm:2003hm, Fayet:2004bw, Nomura:2008ru,
Hochberg:2018rjs, Manzari:2023gkt}, the muon
$g-2$ puzzle~\cite{Muong-2:2023cdq}, the ATOMKI 17 MeV
anomaly~\cite{Krasznahorkay:2018snd, Krasznahorkay:2021joi}, the
$W$-mass tension~\cite{CDF:2022hxs, Bandyopadhyay:2022bgx}, and the
presently subdued---but never disappearing---flavor
anomalies~\cite{Crivellin:2023zui}.

In this note, we will be concentrating on light (\(\sim\) MeV) vector
bosons with vectorial couplings to the standard model (SM) fermions.
Models that belong to this category are varied and multifaceted,
however, owing to the vectorial nature of their couplings, these can be
parametrised in somewhat generalised fashion. The prototypical model
that represents this vast array of models is the dark photon which
interacts with the SM fermions through gauge kinetic mixing with the
photon~\cite{Okun:1982xi, Galison:1983pa, Holdom:1985ag}. The dark
photon (DP), starting out as a theoretical curiosity has gained a central
role in modern particle physics research due to its potential to be a
simple (and predictive) portal to the dark sector of the Universe. In
this paper, we vary the relative coupling strength of the DP between the
visible and the dark sector to define different model benchmarks.

Our aim in this paper is to discuss the sensitivity of the Belle
II~\cite{Belle-II:2018jsg} experiment to the DP parameter space. To be
specific, we are interested in displaced vertex searches that looks for
long lived dark DPs. This choice is informed by the observation that the
region of parameter space that is as of now unexplored lies between
those from recent prompt searches at different
colliders~\cite{BaBar:2014zli, Anastasi:2015qla, NA482:2015wmo,
    BaBar:2016sci, BaBar:2017tiz, LHCb:2017trq, KLOE-2:2018kqf,
    NA62:2019meo, LHCb:2019vmc} sensitive to strong (relatively)
    couplings, and searches at different beam dump
    experiments~\cite{Bjorken:1988as, Riordan:1987aw, Konaka:1986cb,
    Blumlein:1990ay, Blumlein:1991xh, Tsai:2019buq, Davier:1989wz,
APEX:2011dww, CHARM:1985anb, Gninenko:2011uv}, sensitive to very weak
couplings. The region of intermediate coupling strengths lends itself
readily to displaced vertex searches at modern \(\sqrt{s}\sim\) 10 GeV
colliders, constructed primarily to test flavor
observables~\cite{Belle-II:2018jsg, BESIII:2020nme, GlueX:2020idb}.

The results in this paper, in principle new, should be considered as an
addendum to Ref.~\cite{Bandyopadhyay:2022klg}. We employ the same
analysis techniques developed in that paper. In this work, we extend the
results Ref.~\cite{Bandyopadhyay:2022klg} by looking at a different kind
of model that does has couplings, varying from weak to strong, to the
dark sector. As a result of this, not only do our exclusions change, but
the existing constraints also change in nature.  

The paper is organised as follows: in \cref{sec:model} we lay down our
parametrization and compute the relevant decay widths and branching
ratios. In \cref{sec:analysis}, we discuss the analysis techniques
employed to compute the sensitivity of the detector to the displaced
vertices, leading to the 90\% CL exclusions. Then in \cref{sec:results},
we show the exclusions obtained for the different model benchmarks and
discuss the qualitative nature of the contours. Then, in \cref{sec:ex} we
discuss existing exclusions, from prompt and displaced searches. In this
section, for completeness, we also discuss some prospects of displaced
vertex searches at upcoming or existing experiments. Finally, we
conclude.

\section{The Model}\label{sec:model}
Given we are working with a DP (\(A^\prime\)), the couplings to the visible sector are
unambiguously determined by those of the photon~\cite{Okun:1982xi,
Galison:1983pa, Holdom:1985ag}. The DP couples to the SM particles with
a strength that is scaled down---by a factor \(\epsilon\)---from that of
the photon and the scaling is the same for all the SM particles.
However, neither the strength of the DP coupling to the dark sector, nor
the structure of the dark sector are known. To take this into account,
we write a general Lagrangian:
\begin{align}
    \label{eq:lag}
    \mathcal{L}\ &\supset\ \epsilon e \left(\sum_f q_f\bar{f}\gamma^\mu f
    +  \sum_\chi C_\chi\mathcal{O}^\mu_\chi \right)A^\prime_\mu\,.
\end{align}
Here, \(C_\chi\) are the couplings of the DP to dark sector operators
\(O_\chi\). We parametrise the dark sector by ratios of decay width
between the visible and the dark sector. The decay width of the DP to
the visible sector is:
\begin{align}
    \label{eq:visdec}
    \Gamma_V\ &=\ \sum_f \Gamma_f\ =\ \sum_f n^c_f\frac{\epsilon^2 e^2 q_f^2}{12\pi} M_{A^\prime}
    \sqrt{1-4\frac{m_f^2}{{M_A^\prime}^2}} \left( 1+2 \frac{m_f^2}{{M_A^\prime}^2}\right)
\end{align}
Here, $n^c_f$ is the color factor of the fermion $f$. We then take three
model benchmarks, defined in terms of the decay width to the visible
sector, \(\Gamma_V\) and that to the invisible sector, \(\Gamma_I\): i)
\(\Gamma_I=\Gamma_V\), ii) \(\Gamma_I=10\,\Gamma_V\), and iii)
\(\Gamma_I=\Gamma_V/10\). Note, in this way we can effectively
parametrise the dark sector without requiring any specific information
about it. Not, as for this displaced vertex search we only need the
cross section of production of the DP and its branching ratios to
leptons, the analysis is blind to details of the dark sector except the
width.

\begin{figure}[htpb]
    \includegraphics[width=0.47\textwidth]{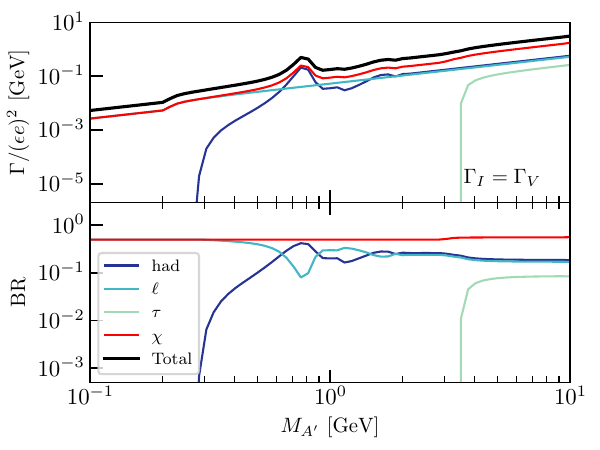}
    \includegraphics[width=0.47\textwidth]{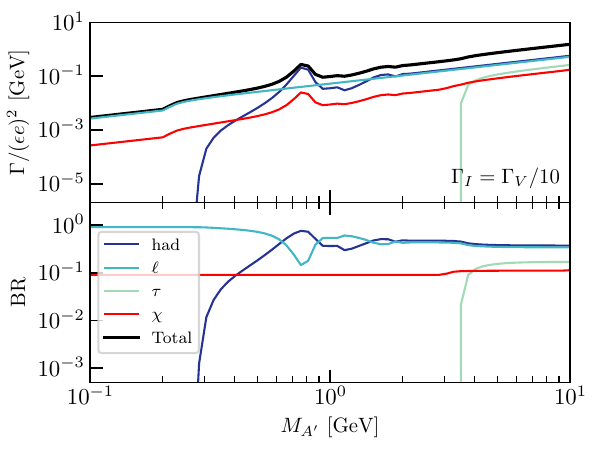}
    \caption{For two different benchmarks, \(\Gamma_I=\Gamma_V\) (left)
    and ii) \(\Gamma_I=\Gamma_V/10\) (right), we plot the decay widths
and branching rations of the dark photon to leptons, hadrons, the tau,
and the dark sector.}
    \label{fig:bw}
\end{figure}

In \cref{fig:bw} we plot the partial widths and branching ratios of the
dark photon to different final states. As we will see later, the
kinematics of the process allow neither the $\tau$ leptons nor the
hadrons to play any part. Note that, the widths as depicted in
\cref{fig:bw} are quite large\footnote{even more so for the
\(\Gamma_I=10\Gamma_V\) case which has not been shown in the figure.}.
However, the decay widths, as plotted, are for the effective coupling,
\(\epsilon e\), taken to be one. For the region that we are interested
in, \(\epsilon e \ll 1\), the decay widths are scaled down well below
the mass of the DP\@. For the decay width of the DP below
\(\Lambda_{\mathrm{QCD}}\), we use the package \texttt{DarkCast}
\cite{Ilten:2018crw} which uses the framework of Vector Meson
Dominance~\cite{Fujiwara:1984mp} to take care of the hadronic
thresholds. 


With the details of our model laid out, we discuss the analysis strategy
to look for the DPs in the next section.
 
\section{The analysis}\label{sec:analysis}
    Our strategy to probe the parameter space of the DP is to look for
    displaced vertex signatures once (if) it is produced. We get a
    displaced vertex is when a particle produced at the interaction
    point (IP) travels a distance away from it before decaying, to
    charged decay products for our case. Tracing the tracks of the decay
    products, the vertex is reconstructed. Following the techniques
    employed in Ref. \cite{Bandyopadhyay:2022klg}, we construct the
    vertex reconstruction efficiency (VRE) distribution as a function of
    the detector parameters and the kinematics of the process. 

    Let us assume that a particle is coming out at an azimuthal angle
    \(\theta\), measured from the \(z\)-axis (along the electron beam).
    If it decays at a distance \(l\) from the IP, then the distances
    along the \(z\) axis and the \(x\)-axis are \(z=l\cos\theta\) and 
    \(r=l\sin\theta\) respectively. If this decay event is within the
    fiducial volume of the tracker, then it is accepted with some
    efficiency. For our purpose, the fiducial volume is defined not by
    the extent of the detector but also by that of efficient track
    reconstruction and background control. 
    
\begin{figure}[htpb]
    \raggedright
    \includegraphics[scale=0.85]{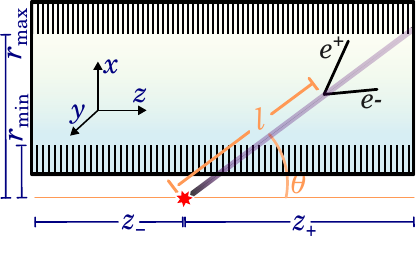}
    \includegraphics[scale=0.85]{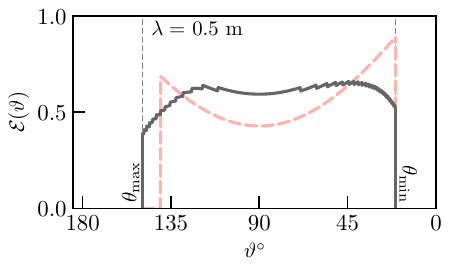}
    \caption{
        \emph{Left:} We plot a schematic of the Belle II tracker. We do
        not distinguish between the silicon inner detectors (VXD) and
        the drift chamber (CDC). We show the radial and axial limits
        within which a decay has to take place for which to accept an
        event. The gradient in the body of the detector represents the
        falling efficiency of detection away from the IP and the
        gradient in the outgoing particle represents the exponentially
        falling decay probability distribution. The coordinate system
        has also been shown. Details can be found in the text.
        \emph{Right:} We show \cite{Bandyopadhyay:2022klg} the Vertex
        Reconstruction Efficiency distribution as a function of the
        scattering angle. In black we show the true distribution and in
        orange we show the distribution we would have obtained if
        instead of convolution with the exponential decay distribution
        we assumed the particle to decay at its characteristic decay
        length ($\lambda=0.5$ cm for this plot.). We also show the cuts
        imposed on the scattering angle. The vertical dashed lines
        (gray) show the cuts imposed on the scattering angle.
}
    \label{fig:car}
\end{figure}

In the left panel of \cref{fig:car} we show a cartoon that shows the
relevant parameters involved. We show a section of the upper half of the
tracking detector. A certain length in the \(x\) direction falls inside
the beam pipe and hence is not accessible. Typically, immediately after
the beam line are the silicon vertex trackers (pixel and strip) with the
highest tracking resolution \cite{Belle-II:2018jsg}. However, this
region is dominated by vertices from long lived SM decays like those of
the \(\Lambda\) and the \(K_L\). There are also tracks from the
secondary interactions of particles with the detector material. Hence,
we look for NP signals farther away from the IP\@, in the central drift
chamber (CDC). This results in the lower cut on the radial distance,
$r_\mathrm{min}$. The upper limit exists to ensure that the DP decays
deep enough inside the tracker for the daughter particles to be
reconstructed. The cuts in the axial length and in the azimuthal angle
comes from the consideration of the detector and beam pipe dimensions.
The \((r,z,\theta)\) triad forms the set of relevant of geometric cuts. 

\begin{table}[htpb]
    \centering
    \caption{Table showing the relevant cuts for our analysis and other
     relevant details of the Belle II experiment (for justifications for
 the cuts, see Ref.~\cite{Bandyopadhyay:2022klg}.)}
    \label{tab:cuts}
    \begin{tabular}{cccccc}
        \toprule
        \multirow{3}{*}{$\sqrt{s}$} & \multirow{2}{*}{Integrated}& \multicolumn{4}{c}{Cuts}   \\
                                   \cmidrule{3-6}
        & Luminosity & radial (cm)& axial (cm)& angular & $p_T$  \\
        \midrule
        10.6 GeV & 50 ab\textsuperscript{-1} & 10 $<r<$ 80 & -40
        $<z<$ 120& $17^\circ<\theta<130^\circ$ & $>$ 0.98 GeV\\
        \bottomrule
    \end{tabular}
\end{table}

For an event that passes the geometric cuts, we look at the efficiency
of reconstruction of the vertex. From the detector side, there is a
linearly falling efficiency (obtained from monte carlo simulations) with
increasing radial distance \cite{Bertholet:2021hjl}:
\begin{align}
    \label{eq:lineff}
    \mathcal{E}(r)\ &=\
    \frac{r_\mathrm{max}-r}{r_\mathrm{max}-r_\mathrm{min}}\,.
\end{align}
In addition, we impose a lower cut on \(p_T\), \(p_T^\mathrm{min}\), as
the vertex reconstruction efficiency drastically falls with very low
momentum These cuts, along with other details of the Belle II detector,
are tabulated in \cref{tab:cuts}.

Given the relevant cuts, we construct the VRE distribution as follows.
After being produced, the dark photon decays at a distance \(l\) from
the IP with a probability of 
\begin{align}
    \label{eq:decpro}
    \mathcal{P}(l)\ &=\ \exp[-l/\lambda]/\lambda\,,
\end{align}
where \(\lambda\) is the characteristic decay length of the dark photon,
given by:
\begin{align}
    \label{eq:chardec}
    \lambda\ &=\ \beta\gamma c\tau\ =\
    \frac{|\vec{p}(M_{A^\prime})|}{M_{A^\prime}} c \tau(M_{A^\prime},
    \epsilon e)\,, 
\end{align}
where $c$ is the velocity of light in vacuum and $\tau(M_{A^\prime},
\epsilon e)$ is the lifetime of the DP as a function of its mass and
coupling strength. Therefore, the efficiency that a DP emitted in the
direction \(\theta\) will be detected at a length \(l\) from the IP is
given by:
\begin{align}
    \begin{split}
    \mathcal{E}_f(\theta)&= \frac{\varepsilon_f}{\mathcal{N}}
        \int_{l_\mathrm{min}}^{l_\mathrm{max}}  \;
         \frac{r_\mathrm{max}-l\, s\theta}{r_\mathrm{max}-r_\mathrm{min}}
        e^{-l/\lambda}\; \overline{\Theta}(l) \; dl\,,
    \\\mathrm{where}&\quad 
        \overline{\Theta}(l)=\Theta\left(l\,s\theta-r_\mathrm{min}\right)\,
        \Theta(r_\mathrm{max}-l\,s \theta)\,
        \Theta(l\,c \theta-z_\mathrm{min})\,
        \Theta(z_\mathrm{max}-l\, c\theta)\;,
    \\\mathrm{and}&\quad l_\mathrm{min}= r_\mathrm{min};\;
    l_\mathrm{max}=\sqrt{r_\mathrm{max}^2 + z_\mathrm{max}^2},\quad
    \mathcal{N}= \int_{l_\mathrm{min}}^{l_\mathrm{max}}\;  
        e^{-l/\lambda}\; \overline{\Theta}(l) \; dl.
    \label{eq:fin_eff}
    \end{split}
\end{align}
Note, with \(s\theta\equiv \sin\theta\), \(c\theta=\cos\theta\), \(l
s\theta\) and \(l c\theta\) give the displacement of the DP in the
radial and the axial direction, respectively. The \(\varepsilon_f\)
factor is an overall detection efficiency of the lepton \(f\) (electron
or muon for us), given by \cite{Belle-II:2018jsg, Bandyopadhyay:2022klg}:
\begin{align}
    \label{eq:oveff}
    \varepsilon_e\ &=\ 0.93\,;\quad \mathrm{and}\quad \varepsilon_\mu\ =\ 0.86\,.
\end{align}
Then, the total number of accepted displaced events over the full run of
Belle II is given by:
\begin{align}
    \label{eq:num_events}
    N_\mathrm{Tot}&= L_\mathrm{I} \times \sum_f
    \int_{c\theta\mathrm{min}}^{c\theta_\mathrm{max}}
    dc\theta\;
    \frac{d\sigma}{dc\theta}\times \mathrm{BR}_{f}\times
    \mathcal{E}_f(\eta)\times 
    \Theta\!\left(p_T^\mathrm{min}-\sin\vartheta(\eta)\,|\vec{p}|\right),
\end{align}
where \(L_I\) is the integrated luminosity,  \(\mathrm{BR}_f\) is the
branching ratio in the \(f\) channel, $\mathcal{E}_f(\eta)$ is the VRE
distribution for the \(f\) channel. The Heaviside function ensures that
the $p_T$ of the DP is greater than $p_T^{\min}$. We are not
including the details of the differential cross-section,
\(\frac{d\sigma}{dc\theta}\), here as the details can be found in
Ref.~\cite{Bandyopadhyay:2022klg}. In the next section, we use the
expression in \cref{eq:num_events} to estimate the reach of the Belle II
detector towards the model benchmarks in question.

\section{Results}\label{sec:results}
For a particular point in the ($M_{A^\prime}, \epsilon e$) plane, we
require the total number of events (\cref{eq:num_events}) to be greater
than three for that point to be accepted. This is because in a virtually
background-free environment, Poisson statistics requires 2.3 events for
exclusion at 90\% CL\footnote{The conservative lower limit on the
    radial distance, \(r>10\) cm, that we allow ourselves effectively
    vetoes SM backgrounds from \(K_L\) and \(\Lambda\) decays and those
from secondary interactions.}. 

\begin{figure}[htpb]
    \centering
    \includegraphics[scale=1]{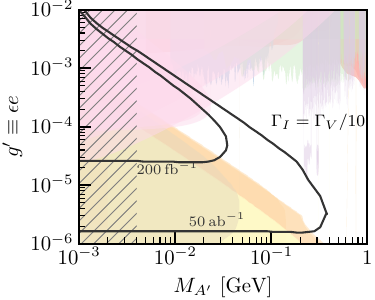}
    \includegraphics[scale=1]{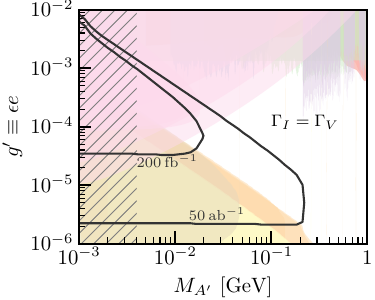}
    \includegraphics[scale=1]{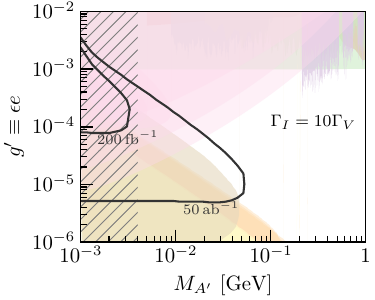}
    \includegraphics[scale=1]{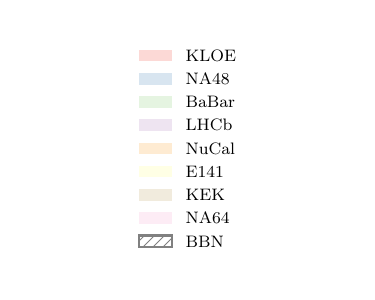}
    \caption{We plot the 90\% CL exclusion contours for different model
        benchmark points, viz., $\Gamma_I=\Gamma_V/10$,
        $\Gamma_I=\Gamma_V$, and \(\Gamma_I=10\Gamma_V\). For
        comparison, we have plotted the exclusion for an integrated
        luminosity of 200~fb\textsuperscript{-1} and
        50~ab\textsuperscript{-1}. As expected, the contours weaken with
        reduction in branching to the visible sector and with falling
        luminosity. Alongside the exclusions obtained by us, we plot the
        exclusions already given by past experiments. In the
        bottom-right panel, we have given a key of the different
        exclusions (details in \cref{sec:ex}).
}\label{fig:cont}
\end{figure}

In \cref{fig:cont} we plot the contours for the three cases discussed in
\cref{sec:model}, viz., $\Gamma_I=\Gamma_V/10$, $\Gamma_I=\Gamma_V$, and
$\Gamma_I=10\Gamma_V$. We also show a comparison between the exclusions
at different integrated luminosities, the projected
50~ab\textsuperscript{-1} and the current ($\sim$)
200~fb\textsuperscript{-1}. From the figures it is clear that when the
branching to the visible sector is substantial, Belle II is sensitive to
a large part of the ($M_{A^\prime},\epsilon e$) parameter space below
500 MeV, with effective couplings as small as $10^{-6}$ being probed.
Note that the upper bound on the sensitivity is somewhat robust against
change of branching ratios. However, the lower limit changes drastically
with varying branching ratios. 

The hatched region to the far left of all the plots marks the bound for
electrophilic light particles from BBN and CMB
observables~\cite{Sabti:2019mhn}. The other shaded regions in the plots
are existing exclusions on the parameter space of the DP\@. We will give
short descriptions of these limits below. The important point to pick up
from the other limits is that the projections we obtain for Belle II
both supplement and complement the existing limits. We will now briefly
discuss the exiting bounds for completeness and reference.

\section{Existing bounds and future prospects}\label{sec:ex}

In \cref{fig:cont}, we have plotted the contours obtained from our
analysis. We have placed the contours on top of existing bounds. In this
subsection, we discuss in brief the sources of these bounds. The
existing bounds can be categorised as bounds from prompt searches which
mostly come from detectors around interaction points, and as bounds from
displaced searches, which almost exclusively come from beam dump
experiments. Not surprisingly, the exclusions we obtain are mostly
overlap the beam dump bounds.  

\subsection{Prompt}
Prompt bounds come both from lepton colliders and hadron colliders. The
most stringent bounds from lepton colliders are those from the BaBar
experiment~\cite{BaBar:2014zli, BaBar:2016sci, BaBar:2017tiz}. These
bounds arise from the processes \(e^+ e^-\to A^\prime \gamma\) and \(e^+
e^-\to \mu^+ \mu^- A^\prime\) and together are sensitive for $20\,
\mathrm{MeV} <M_{A^\prime}< 10\, \mathrm{GeV}$ and constrain till about
$g^\prime>10^{-3}$. In \cref{fig:cont}, the green shaded region
represents the bounds from BaBar. The KLOE experiment also looked for
DPs in the same channel and found a bound, but for smaller windows than
BaBar~\cite{Anastasi:2015qla, KLOE-2:2018kqf}. In \cref{fig:cont}, the
KLOE bounds are given in red.

For the mass range we are probing, \(\pi^0/\eta\to A^\prime \gamma\) and
mixings with the \(\rho, \omega, \phi\) mesons are the primary modes of
production of the DP in hadron colliders. Among the hadron colliders,
the LHCb collaboration has obtained the strongest bounds on the DP
parameter space~\cite{LHCb:2017trq, LHCb:2019vmc}. In \cref{fig:cont},
the violet shade gives the bound from the LHCb collaboration. The CMS
collaboration has also presented bounds on the DP parameter space,
however, the experiment loses sensitivity for DP masses below \(\sim\)
10 GeV \cite{CMS-PAS-EXO-23-014}. Recently, in Ref.
\cite{Bhattacherjee:2023evs} it has been pointed out how machine
learning techniques can be implemented to increase signal efficiency (in
the context of the HL-LHC).
There are bounds on the DP parameter space given by the ATLAS
collaboration as well \cite{ATLAS:2022izj}. However, the exclusions are given in the context
of UV models. We have to reparametrise the analysis to place the bound
on top of our result. The NA48/2~\cite{NA482:2015wmo} and the
NA62~\cite{NA62:2019meo} experiments have looked for dark photons in the
decay products of pions in prompt searches. In \cref{fig:cont} we have
shown the NA48/2 bounds in blue. 

\subsection{Displaced}

To begin with, we have the NA64 experiment using the SPS beam at CERN\@.
In this experiment, a 100 GeV electron beam is focussed on an active
material that acts as a beam dump. The electrons (beam) lose their
energy by emitting bremsstrahlung photons \emph{and}, if present, dark
photons. The DPs being weakly interacting and long lived can escape the
dump. These DPs then decay to a pair of electrons and the energy of the
decay products are measured in an ECAL\@. 

A candidate $A^\prime \to e^+e^-$ event is one where there are
simultaneous showers in the dump and the ECAL, the two adding up to the
beam energy. A mismatch signals channeling of energy into invisibles,
leading to the bound. The bounds shaded in pink in \cref{fig:cont} are
those obtained from the NA64 experiment. As the search strategy is based
on kinematic imbalance, the NA64 bound is highly sensitive to decays
into the dark sector. Hence, the exclusion becomes stronger with
increasing branching to the dark sector, as can be seen from
\cref{fig:cont}.

Other searches at E137~\cite{Bjorken:1988as},
E141~\cite{Riordan:1987aw}, KEK~\cite{Konaka:1986cb},
NuCal~\cite{Blumlein:1990ay, Blumlein:1991xh, Tsai:2019buq},
Orsay~\cite{Davier:1989wz}, APEX~\cite{APEX:2011dww} also work on more
or less the same setup and principles, while for
CHARM~\cite{CHARM:1985anb} and NOMAD~\cite{Gninenko:2011uv}, the
incident particles are protons in place of a hadrons. In
\cref{fig:cont}, we have shown the exclusion contours for NuCAL
(orange), E141 (brown) and KEK (yellow). We have suppressed the other
exclusions as they end up constraining the same region as these three. 

There is a constraint from displaced searches given by the LHCb
collaboration as well~\cite{LHCb:2017trq, LHCb:2019vmc}. As of now, this
region is quite small and vanishes in all cases with substantial
coupling to the dark sector. In \cref{fig:cont} this constraint is
visible as an `island' in the $\Gamma_I=\Gamma_V/10$ case only (violet).

Before ending this section, we would like to mention that alongside
Ref.~\cite{Bandyopadhyay:2022klg}, Ref.~\cite{Ferber:2022ewf} also came
out, where the Belle II projections were somewhat different from those
in Ref.~\cite{Bandyopadhyay:2022klg}. We couldn't plot their contours
side by side as they obtained the constraints for dark matter coupling
going to zero. Qualitatively, their results indicate that Belle II is
sensitive to DP masses much greater than what we get. On the other hand,
their results indicate weaker sensitivity to the coupling strength than
what we get. 

\subsection{Prospects and other analyses}
It is clear that the reason that we can probe such a large region of
parameter space in Belle II is because of the substantial volume of the
CDC around the IP\@. Now, another experiment that has similar tracking
abilities around the IP is GlueX at Jlab~\cite{GlueX:2020idb}. In this
experiment, Bremsstrahlung photons of energy 3-12 GeV (6--22 GeV in the
near future~\cite{Accardi:2023chb}) are incident on
hydrogen nuclei (or heavy ions). Integrated luminosity is expected to
reach above \(\sim\) 1~fb\textsuperscript{-1}. Like Belle II, GlueX has 
a central drift chamber around the interaction point, the details of
which are given in \cref{tab:gx}. 

\begin{table}[htpb]
    \centering
    \caption{Some details of the GlueX experiment.}
    \label{tab:gx}
    \begin{tabular}{ccccc}
        \toprule
        \multirow{3}{*}{$\sqrt{s}$} & \multirow{2}{*}{Integrated}&
        \multicolumn{3}{c}{Active CDC Dimensions}   \\
                                   \cmidrule{3-5}
        & Luminosity & radial (cm)& axial (cm)& angular\\
        \midrule
        3.6 GeV -- 7 GeV& few fb\textsuperscript{-1} & 10 $<r<$ 50 & -50
        $<z<$ 100& $20^\circ<\theta<132^\circ$\\
        \bottomrule
    \end{tabular}
\end{table}

The CoM energy in \cref{tab:gx} assumes that the target is proton, it
will increase for heavy ions. Our initial analysis for GlueX shows that
it is almost as sensitive to the displaced vertex signatures of the DP
as is Belle II\@. The slight disadvantage it has is due to its lower CoM
energy compared to Belle II\@. We will soon make our findings public. 
Note, as GlueX employs fixed target collisions, the final state
particles has significant forward activity. To register the tracks due
to these particles, the experiment has installed a forward tracker with
substantial fiducial volume. We are yet to analyse the sensitivity of
this tracker to DP displaced vertices and hence, have not included
details of this part in \cref{tab:gx}.  

Not only GlueX, the FASER detector at LHC has recently started
publishing their DP results and although, as of now, these bounds are
not much stronger than existing ones, with more data taking it is bound
to cover vast regions of the uncharted parameter
space~\cite{FASER:2023tle}. Other LHC annex experiments like
MATHUSLA~\cite{Curtin:2018mvb} and CODEX-b~\cite{Gligorov:2017nwh} have
also been commissioned/proposed to look for long lived particles and
will no doubt expand on the parameter space probed as of now. Dark
Photons has also been considered in the context of proposed experiments,
like the FCC-hh~\cite{Bhattacherjee:2023plj}, at the
EIC~\cite{Davoudiasl:2023pkq}, and the ILC (and FCC-ee,
CEPC)~\cite{Schafer:2022shi, Asai:2021ehn}.

\section{Conclusions}
The region of unexplored parameter space for a dark photon like particle
lends itself well to displaced vertex searches. With this
identification, we have discussed the sensitivity of the Belle II CDC
towards such vertices. To do so, we have assumed models where the DP is
a portal between the visible and the dark sector and have studied
different scenarios parametrised by the relative widths of the dark
photon to the visible and the invisible sectors. We find the sensitivity
of Belle II to complement existing bounds for large regions of the
parameter space. Our positive findings behove the Belle II collaboration
to analyse their data for displaced vertices of dark photons decaying to
charged leptons. We concluded that with Belle II alongside a plethora of
other experiments equipped to reconstruct displaced vertices, we are
certain to probe extended regions of the unexplored dark photon
parameter space in the current decade.

\bmhead{Acknowledgments}
The author acknowledges Sabyasachi Chakraborty and Sokratis Trifinopoulos for their contributions to the 
development of the analysis techniques and Igal Jaegle for pointing out the suitability of the GlueX 
experiment.

\section*{Declaration}
\textbf{Availability of data and materials:}
Data sets generated during the current study are available from the corresponding author upon reasonable request.


\bibliography{sn-bibliography.bib}

\end{document}